\documentclass[sn-mathphys]{sn-jnl}

\usepackage{amssymb} 
\usepackage{pifont}  
\usepackage{caption} 
\usepackage{subcaption}
\usepackage{multirow}
\usepackage{amsmath,amssymb,amsfonts}
\usepackage{amsthm}
\usepackage{mathrsfs}
\usepackage[title]{appendix}
\usepackage{xcolor}
\usepackage{textcomp}
\usepackage{manyfoot}
\usepackage{booktabs}
\usepackage{algorithm}
\usepackage{algorithmicx}
\usepackage{algpseudocode}
\usepackage{listings}
\usepackage{float}
\usepackage{doi}
\usepackage{url}
\usepackage{placeins}

\theoremstyle{thmstyleone}

\theoremstyle{thmstyletwo}

\theoremstyle{thmstylethree}%

\begin{document}

\title[Article Title]{Adversarial Machine Learning for Robust Password Strength Estimation}

\author{\fnm{Pappu} \sur{Jha}}\email{pappu.jha@usm.edu}
\author{\fnm{Hanzla} \sur{Hamid}}\email{hanzla.hamid@usm.edu}
\author{\fnm{Oluseyi} \sur{Olukola}}\email{oluseyi.olukola@usm.edu}
\author{\fnm{Ashim} \sur{Dahal}}\email{ashim.dahal@usm.edu}
\author{\fnm{Nick} \sur{Rahimi}}\email{nick.rahimi@usm.edu}

\affil{\orgdiv{School of Computing Sciences and Computer Engineering}, 
\orgname{The University of Southern Mississippi}, 
\orgaddress{\street{118 College Drive}, \city{Hattiesburg}, 
\postcode{39406}, \state{MS}, \country{USA}}}

\abstract{Passwords remain one of the most common methods for securing sensitive data in the digital age. However, weak password choices continue to pose significant risks to data security and privacy. This study aims to solve the problem by focusing on developing robust password strength estimation models using adversarial machine learning, a technique that trains models on intentionally crafted deceptive passwords to expose and address vulnerabilities posed by such passwords. We apply five classification algorithms and use a dataset with more than 670,000 samples of adversarial passwords to train the models. Results demonstrate that adversarial training improves password strength classification accuracy by up to 20\% compared to traditional machine learning models. It highlights the importance of integrating adversarial machine learning into security systems to enhance their robustness against modern adaptive threats.}


\maketitle

\section{Introduction}\label{sec1}

Data security is an important endeavor in the current era of digitization. As internet-based technologies become increasingly accessible to the public, people must find a source for securing their information on the internet to protect privacy, security, and confidentiality. Some data, like bank accounts, credit card details, social security numbers,  etc., are so sensitive that if exposed to unauthorized parties, they can lead a person to extreme vulnerabilities, financial loss, and a decline in credibility. There are a number of methods of maintaining data security on Internet platforms. They include biometrics, passkeys, facial recognition, etc. However, they are all complex in nature, so not every commoner feels comfortable using them. Hence, the role of one of the most popular means of security, which is widely used due to its simplicity, becomes significant. This measure is none other than passwords.

Passwords refer to the combination of alphabetical letters (upper/lower cases), numbers, and special characters that are used to verify the authenticity of users before they are granted access to the system on the internet. As people usually have accounts on several websites, they keep one or a few passwords for all of them. Similarly, people tend to use memorable words as their passwords for ease, such as birth dates, names of family members, places, and commonly used phrases. Although these practices are convenient for users, they make passwords susceptible to detection by intruders, allowing unauthorized individuals to access users' accounts. Having so is a gross violation of individuals’ digital privacy. It is therefore essential that strong and varying passwords are used by users on different websites. Given that not all users have expertise in technology, it is the responsibility of the website owner to make them aware of the strength while passwords are being set up.

There are many ways to understand the strength of passwords used in the system. Traditionally, it was possible to determine the strength of passwords by analyzing their length and the nature of their characters. For instance, if a password has many characters, a combination of upper and lower cases, and special characters, then it can be considered strong. However, such visual categorization is not always correct and cannot be considered completely reliable.

Currently, there are several online tools that can check the strength of passwords using lexical rules. Some examples are Password Meter, Microsoft Password Checker, and Google Password Meter \cite{techniques}. Although they are easy to use and access, they are based on a static approach. That is, they cannot evolve to meet the changing patterns of cyberattacks. Additionally, there has been an increasing surge of adversarial passwords that are susceptible to attacks. In simple terms, an adversarial attack refers to a technique of manipulating a model with specially crafted input data for deceptive purposes. Likewise, adversarial passwords are deliberately designed to trick algorithms, causing a discrepancy between their actual strength and the strength assessed by a model. For example, if 'password' is determined as weak, 'p@ssword' will be classified as strong due to the presence of a special character ‘@’. It is, therefore, important to have password-strength checking tools that can accurately predict the strength of passwords without falling into the trap of adversarial password attacks.

To address the aforementioned limitations of password-strength checkers, machine learning-based methods come into play. Using machine learning, it is possible to develop classification models that categorize passwords based on their strengths into the required number of classes with high accuracy. Numerous algorithms can be used for this purpose. 

In this study, we contribute to the research topic in the following ways:
\begin{itemize}
\item We independently collected datasets containing a mixture of adversarial and normal passwords from Kaggle --- an online source.

\item We applied machine learning algorithms, including Random Forest, Logistic Regression, Naive Bayes, Decision Tree, and XGBoost, to develop state-of-the-art classification models that can accurately classify deceptive inputs without being manipulated.

\item We tested the developed models with custom adversarial passwords, which were predicted correctly with high accuracy.
\end{itemize}

This paper is structured as follows: Section II provides a comprehensive overview of existing papers on password strength estimation and adversarial attacks. Section III explains the methodology, including data collection, preprocessing, feature extraction, machine learning models employed, and evaluation of models. Section IV presents the results and discusses the performance of the models. Finally, Section V concludes the study with key findings and suggestions for future research directions.

\section{Literature Review}\label{sec2}

Password strength classification has been widely researched using machine learning techniques. However, while many studies achieve high accuracy, few critically examine their own methodological limitations or address vulnerabilities from deceptive password inputs. This review organizes the literature by methodological approach, evaluates each paper’s strengths and limitations, and identifies gaps in these papers that this study addresses.

\subsection{Traditional Classifiers and Feature-Based Approaches}

Suganya et al. (2010) proposed an early solution, where Support Vector Machines (SVM) were used to classify passwords, and standard filters removed passwords that were too similar to the user’s username or commonly found in dictionaries. The model required a training time of only 10.24 seconds and achieved an accuracy of 98.3\%. However, its approach depended heavily on pre-filtering and fixed rules, which limited its ability to generalize to more nuanced or obfuscated password patterns \cite{filters}.

Asaduzzaman et al. (2024) proposed a lightweight method using Term Frequency-Inverse Document Frequency (TF-IDF) and logistic regression. Their model reached 81\% accuracy when applied to real-world leaked passwords. The method was efficient, but it failed to make use of deeper password features (e.g., entropy, substitutions), which are vital for recognizing deceptive patterns that imitate stronger passwords \cite{asaduzzaman}.

\subsection{Multi-Model Comparisons with Large Datasets}

Sarkar et al. (2022) applied a number of simple and complex algorithms, including Logistic Regression, Decision Tree, Random Forest, Naive Bayes, XGBoost, Support Vector Machine, and Multilayer Perceptron. They have a dataset sample of 80,000 generic passwords, which is a noteworthy value but not significant enough to claim their models as novel. Also, their dataset is composed of 12.35\% of strong passwords, 74.29\% medium, and 13.36\% weak. As the dataset is neither fairly distributed, nor balanced by applying techniques like SMOTE, there is a high probability of biases in their models. Lastly, one of their approaches has achieved an exceptional accuracy of 99\%. They have not proved that it is not an outcome of model overfitting by performing k-fold cross-validation \cite{techniques}.

On the other hand, Rehman et al. (2024) used a dataset of around 700,000 passwords from 000WebHost leak, which makes their models trustworthy at first glance. Likewise, they have applied the TF-IDF method to assign weights to different terms based on their frequency, in order to ensure a balanced dataset across all strength categories. Like Sarkar et al., they have also utilized different algorithms of varied complexities, such as Decision Tree, Logistic Regression, Naive Bayes, Random Forest, XGBoost, Support Vector Machine, and Artificial Neural Network to train their models. In addition, they have optimized hyperparameters to enhance the performance of models. Lastly, they have clearly explained the trade-off between complex algorithms with higher accuracy but longer training times, and simpler ones with lower accuracy but faster training. While their work in password classification appears to be significant, it should be remembered that they have used generic passwords for training, not adversarial, which is a growing concern in password security \cite{methods}.

\subsection{Expanded Classification and Real-World Evaluation}

Vijaya et al. (2024) categorized passwords into five levels (Very Weak to Very Strong) using four classifiers: Decision Tree, MLP, Naive Bayes, and SVM. The best performance was recorded by SVM at 98.3\%. Their use of structured features and a custom taxonomy made it possible to observe strengths in more detail. However, their synthetic dataset did not reflect real-world user behavior, so its usefulness was limited beyond their own scenario \cite{vijaya}.

Comparatively, Asaduzzaman et al. used real-world data to make the information more relevant, but they built simpler forms of the models. This contrast illustrates the trade-off between the depth of features and the practicality of deployment. Neither study explored the effects of intentionally deceptive inputs - a gap this paper addresses \cite{asaduzzaman}.

\begin{table}[!htbp]
\centering
\caption{Analysis of Previous Research Work}
\begin{tabular}{|c|c|c|c|c|c|c|c|c|c|c|}
\hline
\textbf{Study} & \textbf{RF} & \textbf{LOR} & \textbf{NB} & \textbf{DT} & \textbf{XGB} & \textbf{SVM} & \textbf{MLP} & \textbf{LIR} & \textbf{GB} & \textbf{ANN} \\
\hline

{[1]} & \ding{55} & \ding{55} & \ding{55} & \ding{55} & \ding{55} & \checkmark & \ding{55} & \ding{55} & \ding{55} & \ding{55} \\
\hline
{[2]} & \ding{55} & \checkmark & \ding{55} & \ding{55} & \ding{55} & \ding{55} & \ding{55} & \ding{55} & \ding{55} & \ding{55} \\
\hline
{[3]} & \checkmark & \checkmark & \checkmark & \checkmark & \checkmark & \checkmark & \checkmark & \ding{55} & \ding{55} & \ding{55} \\
\hline
{[4]} & \checkmark & \checkmark & \checkmark & \checkmark & \checkmark & \checkmark & \ding{55} & \ding{55} & \ding{55} & \checkmark \\
\hline
{[5]} & \ding{55} & \ding{55} & \checkmark & \checkmark & \ding{55} & \checkmark & \checkmark & \ding{55} & \ding{55} & \ding{55} \\
\hline

\end{tabular}

\vspace{1em}

\caption*{\textit{Abbreviations:} RF: Random Forest, LOR: Logistic Regression, NB: Naive Bayes, DT: Decision Tree, XGB: Extreme Gradient Boosting, SVM: Support Vector Machine, MLP: Multilayer Perceptron, LIR: Linear Regression, GB: Gradient Boost, ANN: Artificial Neural Network}
\end{table}

\subsection{Summary of Gaps and Study Motivation}

Across these studies, accuracy is often prioritized over resilience. Models perform well on clean or synthetic datasets, but are not tested against realistic adversarial conditions like character substitutions (e.g., “p@ssword” instead of “password”) or structural tricks that mislead strength meters. Additionally, few models undergo in-depth evaluation with learning curves or cross-validation, which are essential for ensuring robustness.

This study addresses these gaps by:
\begin{itemize}
    \item Utilizing large, mixed-source datasets containing natural adversarial characteristics.
    \item Preprocessing data to simulate real-world password complexity.
    \item Applying class balancing and 5-fold cross-validation to validate generalizability.
    \item Evaluating models with precision, recall, and F1-score across varied inputs.
\end{itemize}

Through these methods, the study not only builds on past work but extends it to a more security-aware and realistic framework.

\section{Methods}

Our methodology uses supervised machine learning algorithms to develop models that classify passwords based on their strengths. It incorporates four major steps. They include data collection and preprocessing, feature extraction, model training and testing, and evaluation.

\subsection{Data collection and preprocessing}\label{AA}
We collected two datasets from Kaggle that have passwords and their corresponding strengths. The first one is titled “Password Strength Classifier Dataset” by Bhavik Bansal, having more than 669,000 passwords, in which passwords’ strengths are categorized into 0 (weak), 1 (medium), and 2 (strong) classes.

Likewise, the second one is under the title of “Password Strength and Vulnerability Dataset” by Utkarsh Singh, having about 500 passwords, in which there are numerous columns, including rank, password, category, time unit, strength, font size, etc. Because we required only password and strength columns, others were dropped from the datasets. Likewise, the strength is rated from 0 to almost 50. Since we needed to group the strengths into three classes only, they were rearranged in a way that strengths from 1 to 4 were labeled as 0, 5 to 8 as 1, and 9 and above as 2.

We cleaned the above two datasets and combined them into one to create a larger and more diverse dataset. We then performed the remaining actions on the combined data. In this dataset, there was a common presence of password properties related to adversarial attacks, namely character substitutions and deceptive complexity. They serve as effective adversarial inputs that can manipulate the ability of classification models to predict correctly. Additionally, there were also borderline instances that simulated real-world adversarial attacks.

Upon collecting data, several preprocessing steps were applied to ensure data consistency and reliability. Firstly, data was cleaned by correcting inconsistencies in class labels (e.g., replacing "week" with "weak"). Secondly, any entries with missing or duplicate values were removed. Likewise, passwords consisting entirely of special characters or non-alphanumeric content were also excluded. Lastly, the values of numerical strengths were divided into three categories—“Weak,” “Medium,” and “Strong”—to simplify classification.

\begin{figure}
\includegraphics[height=8.5cm]{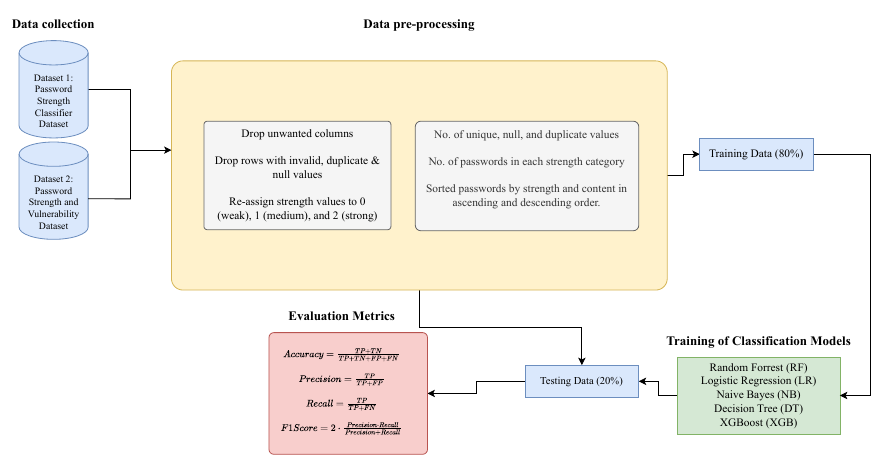}
\caption{Workflow of Password Strength Classification Process}
\label{fig:method}
\end{figure}

\subsection{Feature Extraction}
To extract important features, we examined the datasets focusing on unique values, missing data, and the distribution of passwords across the three strength categories. We also created visualization diagrams, such as heatmaps, to better understand the features in the dataset. Thereafter, we chose the most relevant features—including password length, the number of unique, null, and duplicate values, estimated crack time, and class strength—for model training.

\subsection{Training and Testing the Model}
To develop an effective classification model, several machine-learning algorithms were trained with Python 3.12 using scikit-learn, XGBoost, pandas, matplotlib, joblib, and imbalanced-learn libraries. 

First of all, the dataset was split into 80\% training data and 20\% testing data. To ensure that the splitting of the dataset occurs consistently, a constant value of the random state (e.g., 42) was used. In the same manner, Standard Scaler was applied to normalize numerical features, ensuring consistent input values for the models. It transforms the data so that the mean is equal to 0 and the standard deviation is 1.

\begin{equation}
z = \frac{x - \mu}{\sigma} \label{eq:standardscaler}
\end{equation}

where $z$ is the standardized value (z-score); $x$ is the original value; $\mu$ is the mean of the feature; and $\sigma$ is the standard deviation of the feature.
 
Furthermore, Weights were assigned to classes in a way that minority classes receive higher values compared to the majority. It prevents the minority classes from being underrepresented. For the same purpose, another feature called SMOTE is also applied in some models. It stands for Synthetic Minority Over-sampling Technique. It handles the data imbalance issue by generating synthetic data points for minority classes, which increases their frequency and makes them comparable to the majority.

Moreover, K-Fold Cross Validation was performed to ensure that the overfitting of the data does not occur. It divides the dataset into K folds. Thereafter, a random subset of (K-1) folds is trained and tested against the remaining one fold, and a classification report is generated. The process repeats for K times, and a new test fold is chosen each time. The final classification report is generated by averaging the individual values.

In addition, Learning Curves were generated to understand how our models perform with an increase in training size. The graph contains two curves (training and validation), which represent how the models perform on the respective sets of data. They assist in predicting if the model is underfitting or overfitting.

\subsubsection{Model Selection}
We trained and tested several models using algorithms like Random Forest (RF), Logistic Regression (LR), Naive Bayes (NB), Decision Tree Classifier (DT), and XGBoost Classifier (XGB),

\begin{itemize}
\item Random Forest: It uses multiple decision trees to train a random subset of data separately, make decisions in each tree, and produce the final prediction based on majority voting for classification purposes.

\item Logistic Regression: It is a supervised machine learning algorithm specifically designed for binary classification. However, it can be used for multinomial classification as well. It uses the logistic function to transform the continuous value into a categorical one with the help of a sigmoid function. In simple terms, the sigmoid function is used to map the input variables to a value between 0 and 1. Mathematically,

\begin{equation}
\sigma(z) = \frac{1}{1 + e^{-z}} \label{eq:sigmoid}
\end{equation}

where $z$ is an input to the sigmoid function; $e$ is Euler’s number (the base of the natural logarithm); $\sigma(z)$ is an output probability ranging between 0 and 1; $\sigma(z) \to 1$ as $z \to \infty$; and $\sigma(z) \to 0$ as $z \to -\infty$.

\item Naive Bayes: It is also a supervised machine learning algorithm that performs classification based on the probabilities of classes given the features of the data. It is based on Bayes' Theorem, which is used to determine the conditional probability of an event based on a prior incident. Mathematically, it can be summarized as

\begin{equation}
P(M|N) = \frac{P(N|M) P(M)}{P(N)} \label{eq:bayes}
\end{equation}

where $P(M)$ is the probability of event M; $P(N)$ is the probability of event N; $P(N \mid M)$ is the probability of N given M; and $P(M \mid N)$ is the probability of M given N.

\vspace{0.5em}

\item Decision Tree Classifier: A simple yet powerful model that splits data into decision nodes based on feature importance. It is computationally efficient and interpretable, but it can overfit if not implemented effectively. Decision trees work by identifying the feature that provides the highest information gain and then partitioning the dataset accordingly.

\item XGBoost Classifier with Hyperparameter Tuning: A more advanced gradient boosting algorithm that builds multiple weak learners sequentially to enhance classification accuracy. XGBoost is known for its speed and scalability, incorporating regularization techniques like L1 and L2 penalties to reduce overfitting. The model was fine-tuned using GridSearchCV to optimize hyperparameters such as learning rate, tree depth, and the number of estimators, ensuring better performance.

\end{itemize}

Once each model was trained, it was tested with user-assigned inputs. The generated outputs were then compared with the actual ones to evaluate the model's performance. After testing, all models were saved using Joblib, allowing for future use in password security analysis.

\subsection{Evaluation}
To evaluate the performance of our models, several metrics were used, including Confusion Matrix, Accuracy, Precision, Recall, F1-Score, and Support.

Notably, the Confusion Matrix is an $N \times N$ matrix used in classification to evaluate the performance of a machine learning model. Its components are True Positive (TP), False Positive (FP), True Negative (TN), and False Negative (FN).

Likewise, Accuracy refers to the proportion of all classifications that are correct, both positive and negative. Mathematically, Accuracy = (TP + TN)/(TP + TN + FP + FN).

Precision is the proportion of all the model's positive classifications that are actually positive. Mathematically, Precision = (TP)/(TP + FP).

Further, Recall is the proportion of all actual positives that are correctly classified as positives. Mathematically, it is equal to (TP)/(TP + FN)

$F_1$-score is the harmonic mean of precision and recall. Mathematically, $F_1$ score = (2TP)/(2TP + FP + FN)

Support is equal to the number of actual instances (samples) of a given class present in the dataset used to evaluate the model.

\section{Results and Discussion}

In the study, we worked with a total of five machine learning algorithms meant for classification tasks, including Random Forest, Logistic Regression, Naive Bayes, Decision Tree, and XGBoost. The dataset contained 80\% training data and 20\% testing data. The classification performance was evaluated with Accuracy, together with Precision (Macro), Recall (Macro), and $F_1$-score metrics. The confusion matrix, together with the classification report, generated valuable information about how the classification method reacted to various categories.

\vspace{-1em}

\begin{figure}[H]
    \centering
    \includegraphics[width=0.7\textwidth]{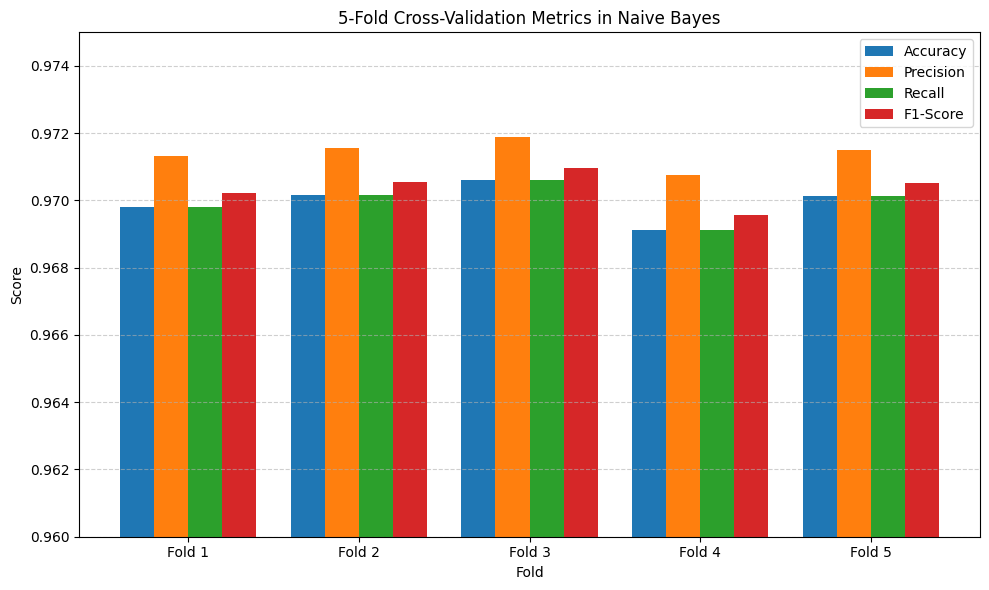}
    \caption{5-Fold Cross-Validation in Naive Bayes}
    \label{fig:cross_validation}
\end{figure}

\vspace{-4em}

\begin{figure}[H]
    \centering
    \includegraphics[width=0.7\textwidth]{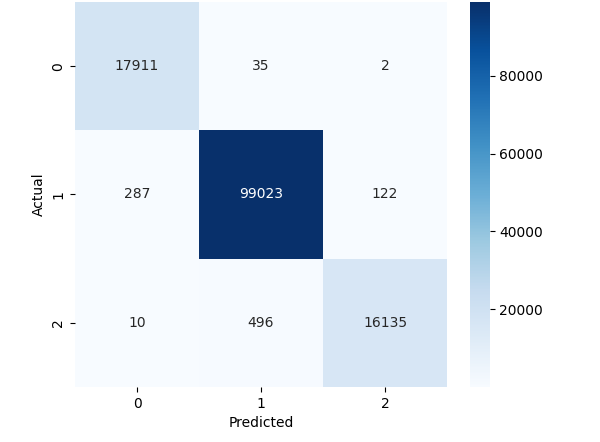}
    \caption{Confusion Matrix of Logistic Regression}
    \label{fig:confusion_matrix}
\end{figure}

\vspace{-4em}  

\begin{figure}[H]
    \centering
    \includegraphics[width=0.7\textwidth]{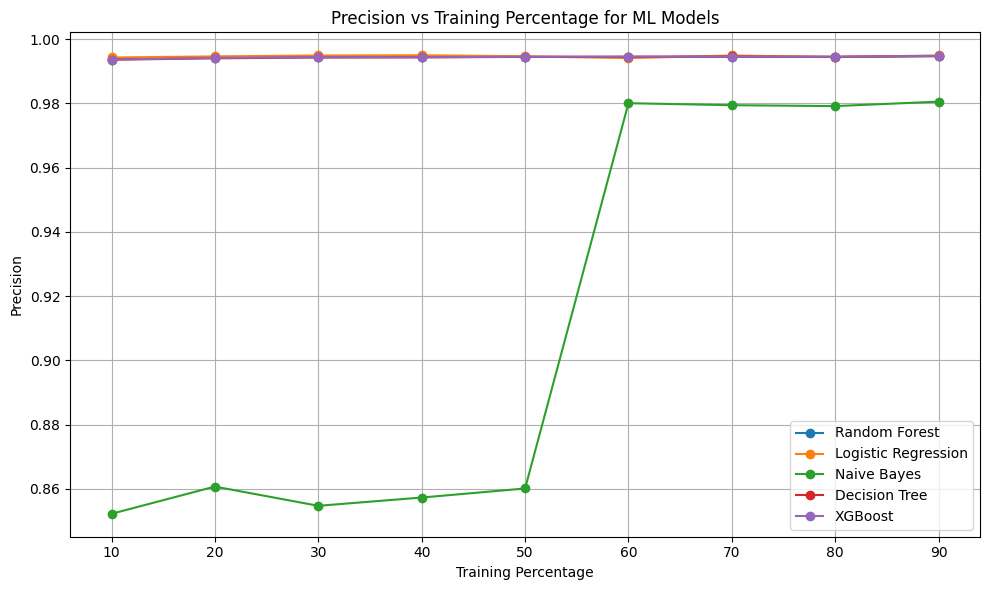}
    \caption{Precision}
    \label{fig:precision}
\end{figure}

\vspace{-4em}  

\begin{figure}[H]
    \centering
    \includegraphics[width=0.7\textwidth]{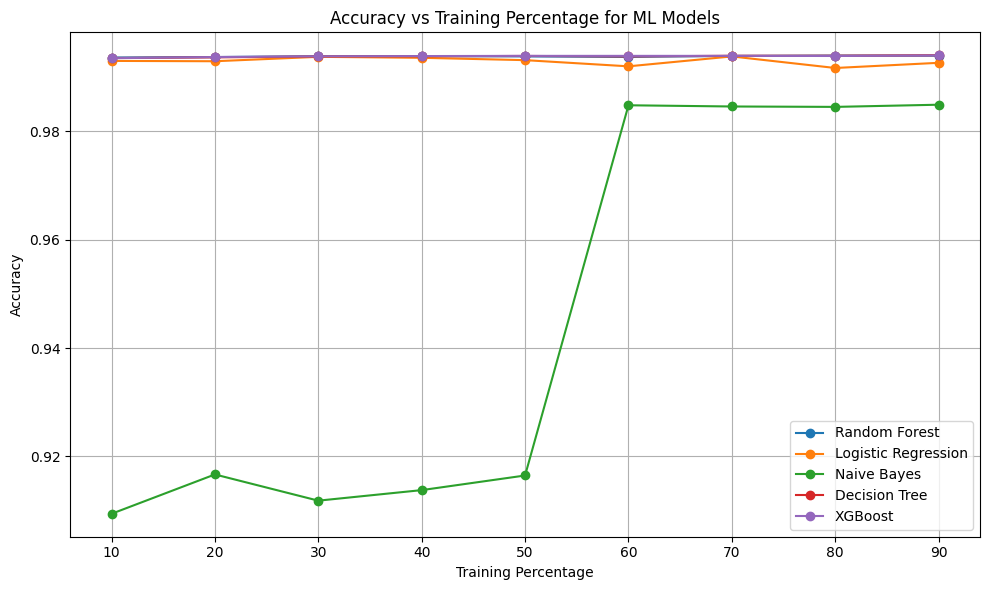}
    \caption{Accuracy}
    \label{fig:accuracy}
\end{figure}

\vspace{-4em}  

\begin{figure}[H]
    \centering
    \includegraphics[width=0.7\textwidth]{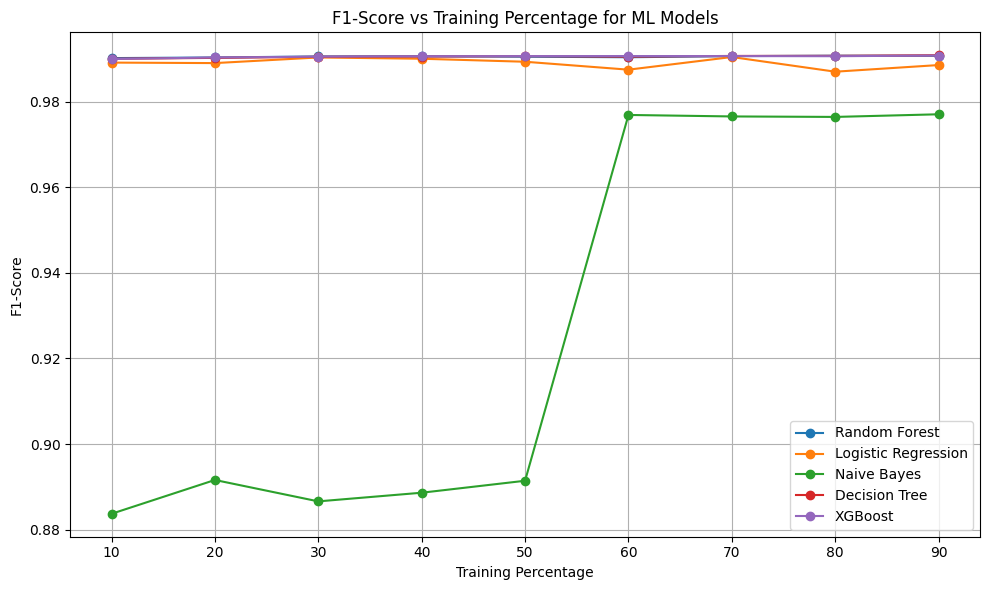}
    \caption{$F_1$-Score}
    \label{fig:f1_score}
\end{figure}

\begin{table}[htbp]
\caption{Results of Our Machine Learning Models}\label{tab:ml_results}
\centering
\begin{tabular*}{\textwidth}{@{\extracolsep{\fill}}lccccc}
\toprule
\textbf{Models} & \textbf{Precision} & \textbf{Recall} & \textbf{$F_1$-Score} & \textbf{Support} & \textbf{Accuracy} \\
\midrule
Random Forest (RF) & 0.99 & 0.99 & 0.99 & 298,295 & 99\% \\
Logistic Regression (LR) & 0.99 & 0.99 & 0.99 & 134,021 & 99\% \\
Naive Bayes (NB) & 0.88 & 0.96 & 0.91 & 201,031 & 94\% \\
XGBoost (XGB) & 0.99 & 0.99 & 0.99 & 134,021 & 99\% \\
Decision Tree (DT) & 0.99 & 0.99 & 0.99 & 134,021 & 99\% \\
\botrule
\end{tabular*}

\footnotesize{Abbreviations: LR = Logistic Regression, RF = Random Forest, NB = Naive Bayes, XGB = XGBoost, DT = Decision Tree}
\label{tab:ml_results}
\end{table}

To ensure that the high accuracy rates of the models are not due to overfitting, we conducted 5-fold cross-validation on all of them. For instance, the results for Naive Bayes are illustrated in Fig.~\ref{fig:cross_validation}. It can be observed that the model consistently performs well across all metrics in each fold of model training and testing, with all values exceeding 0.95.

To understand the $TP$, $FP$, $TN$, and $FN$ values, we generated Confusion Matrices for all models, and the one of Logistic Regression is shown in Fig.~\ref{fig:confusion_matrix}, as an example. For class 0 (weak), 17911 passwords are accurately predicted, and only 297 are false. Likewise, the model is able to successfully classify 99023 and 16135 passwords for class 1 (medium) and class 2 (strong), respectively. The number of falsely predicted passwords for these two classes is negligible compared to the overall size of the dataset.

Furthermore, figures ~\ref{fig:precision}, ~\ref{fig:accuracy}, and ~\ref{fig:f1_score} demonstrate how precision, accuracy, and $F_1$-score perform as the sizes of training data increase from 10\% to 90\%. Across all figures, there is a consistent pattern among models. Random Forest, Decision Tree, and XGBoost achieve the highest performance scores from the beginning, even with the smallest training data, which reflects their robustness. Similarly, Logistic Regression is close to the maximum value and almost reaches it as the training data increases. Lastly, Naive Bayes achieves a low result for all metrics with small training data and grows exponentially when input data crosses 50\%. It tells us that Naive Bayes requires a large amount of data to generalize well.

\section{Conclusion}
In this study, we attempted to develop machine learning models that can detect adversarial passwords and classify them accurately. By reviewing existing research papers on password classification, we identified a lack of studies focused on developing adversarial models for this purpose. Using Kaggle, we found datasets with more than 670,000 samples, which contained several instances of adversarial passwords. Using them to train models, we applied five classification algorithms, including Random Forest, Logistic Regression, Naive Bayes, Decision Tree, and XGBoost, to generate classification models.

Our experiments proved that models developed with adversarial passwords detected adversarial attacks significantly better than traditional models. Notably, our models outperformed existing models by up to 20\% demonstrating the significance of integrating adversarial models in the user authentication system. Remarkably, it will help detect adversarial inputs, protect the data security of users, and avoid cybersecurity breaches on a large scale.

However, a noticeable factor is that the samples of adversarial passwords were naturally present in the datasets, which may not fully capture the essence of the adversarial inputs while training the models. Hence, there is a probability of biases in the models. As part of future work, we will utilize deep learning approaches, including Generative Adversarial Networks (GANs), to generate a more controlled adversarial dataset. After that, we will apply deep learning algorithms like Recurrent Neural Networks (RNNs) and Long Short-Term Memory Networks (LSTMs) to train models and evaluate them with their respective classification reports.

In conclusion, this study demonstrates the significance of adversarial models for successful password classification. As cyberattacks continue to surge and become sophisticated, the need for robust models increases to advance the user authentication system.

\section*{Acknowledgment}
The study was supported by the School of Computing Sciences and Computer Engineering of the University of Southern Mississippi.

\end{document}